\documentstyle[aps,preprint]{revtex}
\def\lsim{\mathrel{\raise2pt\hbox to 8pt{\raise -5pt\hbox{$\sim$}\hss{$<$}}}}

\input BoxedEPS
\SetepsfEPSFSpecial

\begin{document}

\title{The Resurrection of a Symmetry in Nucleon -
Nucleus Scattering}
\author{ {Joseph N. Ginocchio}} \address {Theoretical Division, MS B283, Los Alamos National Laboratory, Los Alamos, New
Mexico 87545, USA}
\maketitle

\begin{abstract}
We re-examine a symmetry in nucleon - nucleus scattering that previously had been proclaimed to be dead.  We show that this
symmetry is the continuum analog of pseudospin symmetry, a relativistic symmetry which manifests itself in the spectra of
nuclei.  Using experimental data only we show that pseudospin symmetry in nucleon - nucleus scattering is not dead but only
modestly broken for 800 MeV proton scattering on nuclei. 
\end{abstract}
\pacs{24.10.Ht,24.10.Jv,24.80.+y,25.40.Cm}

For nucleons moving in a relativistic mean field with scalar $V_S$ and vector potentials $V_V$,
an SU(2) symmetry exists for the case for which $V_S = - V_V$ \cite {bell}. This symmetry
manifests itself in nuclear spectra as a slightly broken symmetry \cite {gino,gino2,ami,gino3}
since
$|{V_S + V_V
\over V_S - V_V}|$ is small for realistic mean fields \cite {wal,mad,arima,ring} and QCD sum rules \cite {cohen}, and, 
in fact, gives rise
to what has been called ``pseudospin symmetry''.  The original observations that led to the
coining of the word ``pseudospin symmetry'' were quasi-degeneracies in spherical shell model
orbitals with non - relativistic quantum numbers ($n_r$,
$\ell$, $j =
\ell + 1/2)$ and ($n_{r}-1, \ell + 2$, $j = \ell + 3/2$) where $n_r$, $\ell$, and $j$ are 
the single-nucleon radial, orbital, and total 
angular momentum quantum numbers, respectively
\cite {kth,aa}.  This doublet structure is expressed in
terms of a ``pseudo'' orbital angular momentum 
$\tilde{\ell}$ = $\ell$ + 1, the average of the orbital angular momentum of the two 
states in doublet, and ``pseudo'' spin, $\tilde s$ = 1/2.
For example,
$(n_r s_{1/2},(n_r-1) d_{3/2})$ will have
$\tilde{\ell}= 1$ , $(n_r p_{3/2},(n_r-1) f_{5/2})$ will have $\tilde{\ell}= 2$, etc. 
These doublets are almost degenerate with
respect to pseudospin, since $j = \tilde{\ell}\ \pm \tilde s$ for the two states 
in the doublet. 
Pseudospin ``symmetry'' was shown
to exist in deformed nuclei as well \cite {bohr,draayer3} and has been used to explain
features of deformed nuclei, including superdeformation \cite {dudek} and identical bands
\cite {twin,stephens}. However, the origin of pseudospin symmetry remained a mystery and ``no deeper understanding of the origin of
these (approximate) degeneracies'' existed
\cite {ben}. The source of pseudospin symmetry as a broken relativistic symmetry of the Dirac Hamiltonian 
with $V_S \approx - V_V$ was pointed out \cite {gino,gino2,ami,gino3}. For spherical nuclei,
pseudo-orbital angular momentum $\tilde{\ell}$ is also conserved and physically is the ``orbital
angular momentum'' of the lower component of the Dirac wavefunction.

One consequence of this relativistic SU(2) pseudospin symmetry is that the
spatial wavefunction for the lower component of the Dirac wavefunctions will be equal in shape
and magnitude for the two states in the doublet. This has been shown to be approximately valid
for realistic relativistic mean fields
\cite {gino2,ami,gino3,ring}.  Recently this approximate equality has been applied to predicting
relationships between magnetic dipole properties of the nucleus and between Gamow-Teller decays
\cite {gino4}

For nuclear bound states the scalar and vector relativistic potentials are
real. However, this symmetry  exists for complex mean fields as well; that is, for
the scattering of nucleons in the mean field of a nucleus. In fact, proton scattering on nuclei
is very well described by treating the nucleon as a Dirac particle moving in an complex scalar,
$V_S$, and vector, $V_V$, optical potential with $V_S
\approx -V_V$ \cite {dave,bunny}. In a paper entitled ``Sudden Death of a Symmetry'' \cite {fred}, the
authors  used the fact that the Dirac Hamiltonian has a symmetry for $V_S = -V_V$ to
predict the analyzing power and spin rotation function for proton scattering.  Since the
experimental data does not agree with this prediction, they correctly concluded that the symmetry is
broken for nucleon - nucleus scattering. Furthermore, they demonstrated that the pseudospin symmetry
breaking depends on the nucleon energy and may decrease as the nucleon energy
is increased. 

In light of the recent success of this broken pseudospin symmetry in explaining the
bound state properties of nuclei, we revisit the application of broken pseudospin symmetry to
nucleon - nucleus scattering.  First we discuss the conventional formalism of scattering in terms of
spin.  We then discuss the formalism of scattering in terms of pseudospin. Finally we discuss
pseudospin symmetry as a broken symmetry and extract from the experimental data an empirical estimate
of the amount of pseudosymmetry breaking in medium energy nucleon - nucleus scattering.

The scattering amplitude, $f$, for the elastic scattering of a nucleon with momentum $k$ on a spin zero target is given by
\cite {herman}:
\begin{equation}
f = A(k,\theta) + B(k,\theta) {\vec \sigma}\cdot {\vec n} 
\label {f}
\end{equation} 
where ${\vec n} $ is the unit vector perpendicular to the scattering plane and
$\theta$ is the scattering angle, 
\begin{equation}
{\vec k}_i\cdot {\vec k}_f = k^2 cos(\theta), 
\label {angle}
\end{equation} 
$\hbar {\vec k}_i$ is the incident momentum, $\hbar {\vec k}_f$ is final momentum, and $\vec{\sigma}$ are the Pauli spin matrices.
The spin independent scattering amplitude, $A$, and the spin dependent amplitude, $B$, can be
expanded in terms of partial waves,

$$
 A = {-i\over 2\ k}\sum_{\ell} [ ({S(k)}_{{\ell}, {\ell} + 1/2}
-1)
\ ( {\ell} + 1) + ({S(k)}_{{\ell}, {\ell} - 1/2} -1)\ {\ell}]\ P_{{\ell}}(cos(\theta)),
$$
\begin{equation}
B = 
{1\over 2\ k} \sum_{\ell} [{S(k)}_{{\ell}, {\ell} + 1/2} -{S(k)}_{{\ell}, {\ell} - 1/2})\ ]
\ P_{{\ell}}^{(1)}(cos(\theta)).
\label {A,B}
\end{equation}
where ${S(k)}_{{\ell}, j}$ is the partial wave scattering amplitude with orbital angular momentum
$\ell$ and total momentum $j$, $j = \ell \pm 1/2$, and $P_{{\ell}}(cos(\theta))$ and
$P_{{\ell}}^{(1)}(cos(\theta))$ are the Legendre and associated Legendre polynomial of rank
$\ell$, respectively.

The differential cross section is given by the total absolute square of the scattering amplitude
average over the spin,
\begin{equation}
{d\sigma\over d\Omega}(k,\theta) = |A|^2 + |B|^2.
\label {sigma}
\end{equation}
By measuring the asymmetry in the cross section with respect to the spin, the polarization can
be determined,

\begin{equation}
 P(k,\theta) = {{B}\ {A}^* + {B}^*\ {A}\over |{A}|^2 + |{B}|^2},
\label {P}
\end{equation}
and by measuring the asymmetry in the cross section with respect to the spin in a second
scattering the spin rotation function can be determined,
\begin{equation}
Q(k,\theta) = {{i({B}\ {A}^* - {B}^*\ {A})} \over |{A}|^2 + |{B}|^2 } 
\label {Q}
\end{equation}

Clearly, if the scattering function does not depend on spin, ${S}_{{\ell}, {\ell} +
1/2}={S}_{{\ell}, {\ell} - 1/2}$, then $B$ = 0, and both the polarization, $P$, 
and the spin rotation function, $Q$, will both vanish.  In Figure 1 we show examples of
$P$ and $Q$ for 800 MeV proton scattering on $^{208}Pb$ \cite {john}.

As outlined in the Introduction the pseudo-orbital angular momentum related to the orbital
angular momentum as \cite{kth,aa,gino}
\begin{equation}
\tilde {\ell} = \ell + 1, j = \ell + 1/2,
\tilde {\ell} = \ell - 1, j = \ell - 1/2.
\label {ps}
\end{equation}
Therefore we define the scattering amplitudes for partial pseudo-orbital angular momentum as,
\begin{equation}
{\tilde S}_{{\tilde \ell}, j = {\tilde \ell} -1/2} =  S_{{\tilde \ell} - 1, j = {\tilde
\ell} -1/2},\\
{\tilde S}_{{\tilde \ell}, j = {\tilde \ell} + 1/2} =  S_{{\tilde \ell} + 1, j = {\tilde
\ell} + 1/2}.
\label {Sps}
\end{equation}
If we substitute these relations into (\ref{A,B}) and use relationships between the
Legendre polynomials, we find that the pseudo scattering amplitudes,
$$
 {\tilde A} = {-i\over 2\ k}\sum_{\tilde \ell} [ ({\tilde S}_{{\tilde \ell}, {\tilde \ell} + 1/2}
-1)
\ ( {\tilde
\ell} + 1) + ({\tilde S}_{{\tilde \ell}, {\tilde \ell} - 1/2} -1)\ {\tilde \ell}]\ P_{{\tilde
\ell}}(cos(\theta)),
$$
\begin{equation}
{\tilde B} = 
{-1\over 2\ p} \sum_{\tilde \ell} [{\tilde S}_{{\tilde \ell}, {\tilde \ell} + 1/2} -{\tilde
S}_{{\tilde \ell}, {\tilde \ell} - 1/2})\ ]\ P_{{\tilde \ell}}^{(1)}(cos(\theta)).
\label {A,Btilde}
\end{equation}
are related by a unitary transformation,
\begin{equation}
\left ({{\tilde A} \atop {\tilde B}} \right ) = \left ( {cos(\theta) \atop isin(\theta)}
{isin(\theta)
\atop cos(\theta) } \right ) \left ({{A} \atop {B}}\right )
\label {unitary}
\end{equation}

The polarization and the spin rotation function then become:
$$
 {P} = {{\tilde B}\ {\tilde A}^* + {\tilde B}^*\ {\tilde A}\over |{\tilde A}|^2 + |{\tilde B}|^2
} ,
$$
\begin{equation}
Q = {sin(2\theta) [ |{\tilde A}|^2 - |{\tilde B}|^2 ] + i cos(2 \theta)[{\tilde B}\ {\tilde
A}^* - {\tilde B}^*\ {\tilde A}] \over |{\tilde A}|^2 + |{\tilde B}|^2}. 
\label {PQtilde}
\end {equation}
In the pseudospin symmetry limit ${\tilde S}_{{\tilde \ell}, {\tilde \ell} + 1/2} = {\tilde
S}_{{\tilde \ell}, {\tilde \ell} - 1/2}$ and hence ${\tilde B}$ will vanish in this limit.
Therefore, $P = 0$, just as in the spin limit, but $Q = sin \ (2\theta)$ \cite {fred}.

Of course, just as for the nuclear bound states, pseudospin is broken because of the approximate
relation between vector and scalar potentials, $V_S \approx - V_V$.  We shall now extract the
amount of pseudospin breaking from the experimental data on the polarization and spin rotation
function \cite {john}. 

We can write the complex
amplitudes as ${\tilde A} = |{\tilde A}|e^{i\phi_{\tilde A}}, {\tilde B} = |{\tilde
B}|e^{i\phi_{\tilde B}}$. In the pseudospin symmetry limit ${|{\tilde B}|\over |{\tilde A}|} = 0$. Under
the assumption that
${|{\tilde B}|\over |{\tilde A}|}$ is small, we expand $P$ and $Q$ to leading order in
${|{\tilde B}|\over |{\tilde A}|}$:

$$
 {P} \approx 2\ {|{\tilde B}|\over  |{\tilde A}|} cos({\phi}_{{\tilde A}} - {\phi}_{{\tilde B}}),
$$
\begin{equation}
Q \approx sin(2\theta) + 2\ cos(2 \theta) {|{\tilde B}|\over  |{\tilde A}|} sin({\phi}_{{\tilde
A}} - {\phi}_{{\tilde B}}).
\label {psapprox}
\end{equation}
Therefore we can solve for ${|{\tilde B}|\over |{\tilde A}|}$, 
\begin{equation}
\left ({|{\tilde B}|\over |{\tilde A}|}\right)^2 = \left ({P\over 2} \right )^2 + \left ( {{Q -
sin\ (2\theta)}
\over {2\ cos\ (2\theta)}} \right )^2 
\label {B/Aps}
\end{equation}
Using the experimental data in Figure 1, we plot in Figure 2 each of the terms in (\ref {B/Aps})
and, in Figure 3, the total, to give us the ratio $ \left ({|{\tilde B}|\over |{\tilde
A}|}\right)^2$. Since $|{\tilde B}|$ depends on $P_{{\tilde \ell}}^{(1)}$ ( see \ref {A,Btilde}),
$|{\tilde B}|$ must go to zero for $\theta = 0^o$ and $\theta = 180^o$. The ratio $ \left ({|{\tilde B}|\over |{\tilde
A}|}\right)^2$ rises from zero for $\theta \approx 3^o$ to a maximum of 10\% for about $\theta \approx 15^o$; the maximum angle
for which both $P$ and $Q$ are measured is $\theta \approx 17^o$. We notice that the second term in (\ref {B/Aps} ) is very small
and the total is mostly due to $\left ({P\over 2} \right )^2 $. Therefore, from Figure 1, we can see that the maximum
reached for $\left ({|{\tilde B}|\over |{\tilde A}|}\right)^2$ is about 12\% even out to $\theta \approx 32^o$.

For comparison we do a similar analysis for the spin amplitudes. If we assume ${|{B}|\over |{A}|}$ small we
obtain,
$$
P \approx 2\ {|{B}|\over  |{A}|} cos({\phi}_{{A}} - {\phi}_{{B}}),
$$
\begin{equation}
Q \approx  2\ {|{B}|\over  |{A}|} sin({\phi}_{{A}} -
{\phi}_{{B}}),
\label {sapprox}
\end{equation}
from which we solve for $ \left ({|{ B}|\over |{A}|}\right)^2$,
\begin{equation}
\left ({|{B}|\over |{ A}|}\right)^2 \approx \left ({P\over 2} \right )^2 + \left ( {Q 
\over 2} \right )^2. 
\label {B/A}
\end{equation}
In Figure 2 we plot each of the terms in (\ref {B/A}).  We notice that the first term is the same as the first term in 
pseudospin breaking but the second term is much larger than the second term for pseudospin, even larger than first term,
leading to a spin breaking magnitude $\left ({|{B}|\over |{ A}|}\right)^2$ of about 22\% even within $\theta \approx 17^o$
(Figure 3), more than twice the pseudospin breaking magnitude $\left ({|{\tilde B}|\over |{\tilde A}|}\right)^2$.

In summary, we have determined directly from the experimental data \cite {john} the pseudospin
breaking for 800 MeV proton scattering from
$^{208}Pb$.  We find that, up to the angles measured (about a momentum transfer of 2.3 fm$^{-1}$), the pseudospin symmetry
breaking depends on the scattering angle but reaches a maximum of 10\% whereas the spin breaking reaches a maximum of 22\%.
Hence, pseudospin has validity for medium energy nucleon scattering. In the future the energy
dependence of pseudospin breaking shall be determined. Hopefully more data will be forthcoming for
the very low nucleon energies for which pseudospin symmetry breaking may be much larger \cite
{fred}.     

The author would like to thank D. G. Madland for discussions and for a compilation of the
experimental data. This research is supported by the U. S. Department of Energy
under contract W-7405-ENG-36.\\

\begin{figure}
\caption{800 MeV Proton Scattering on $^{208}Pb$ versus the scattering angle. The polarization, P, is the solid line; the spin
rotation function, Q, is the dashed line. The lines are a guide to the eye. }
\label{1}
\end{figure}
\begin{figure}
\caption{$\left({P\over 2}\right)^2$ (solid line), $\left( {Q \over 2} \right)^2$ (dashed line), and $\left[{(Q - sin\ (2\theta))
\over (2\ cos\ (2\theta))}\right]^2$ (dotted line) versus the scattering angle. The lines are a
guide to the eye. }
\label{2}
\end{figure}
\begin{figure}
\caption{The spin breaking $\left({|B|\over |A|}\right)^2$ (solid line) and pseudospin breaking $\left({|{\tilde B}|\over|{\tilde
A}|}\right)^2$ (dashed line) versus the scattering angle. The lines are a guide to the eye.}
\label{3}
\end{figure}
\pagebreak
\HideDisplacementBoxes
\hskip1.0truein 
\BoxedEPSF{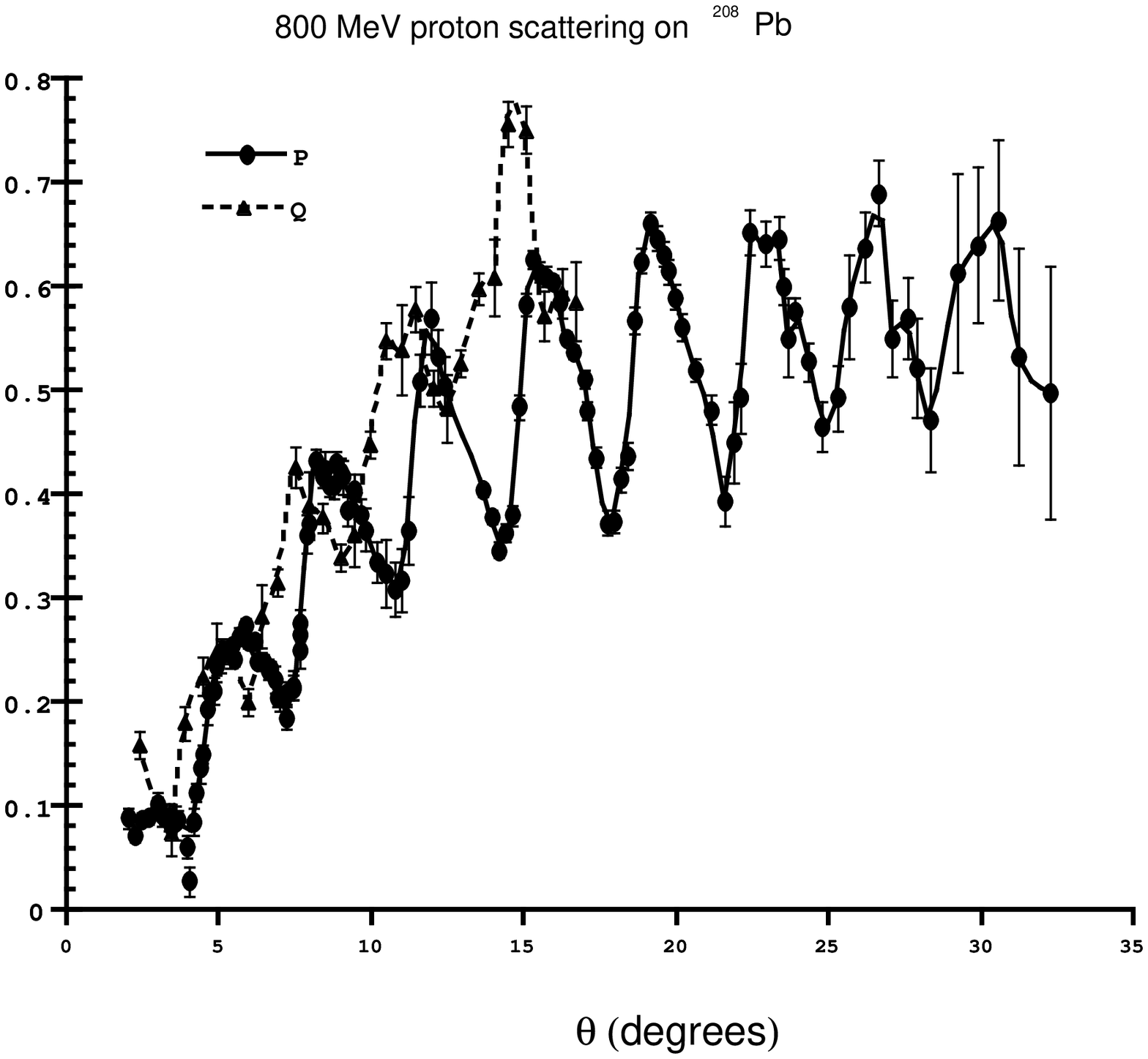 scaled 800}

\HideDisplacementBoxes
\hskip1.0truein 
\BoxedEPSF{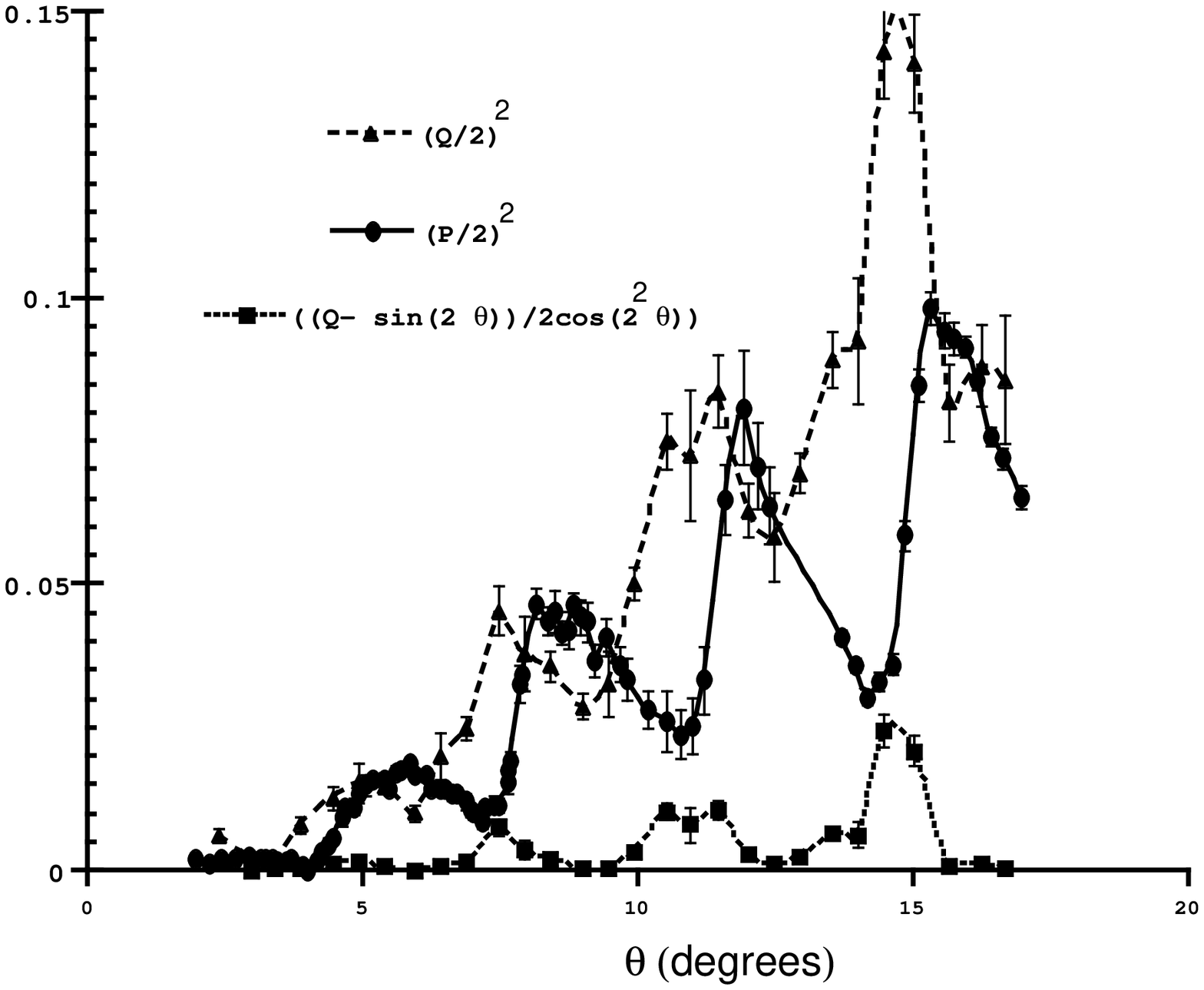 scaled 800}

\HideDisplacementBoxes
\hskip1.0truein 
\BoxedEPSF{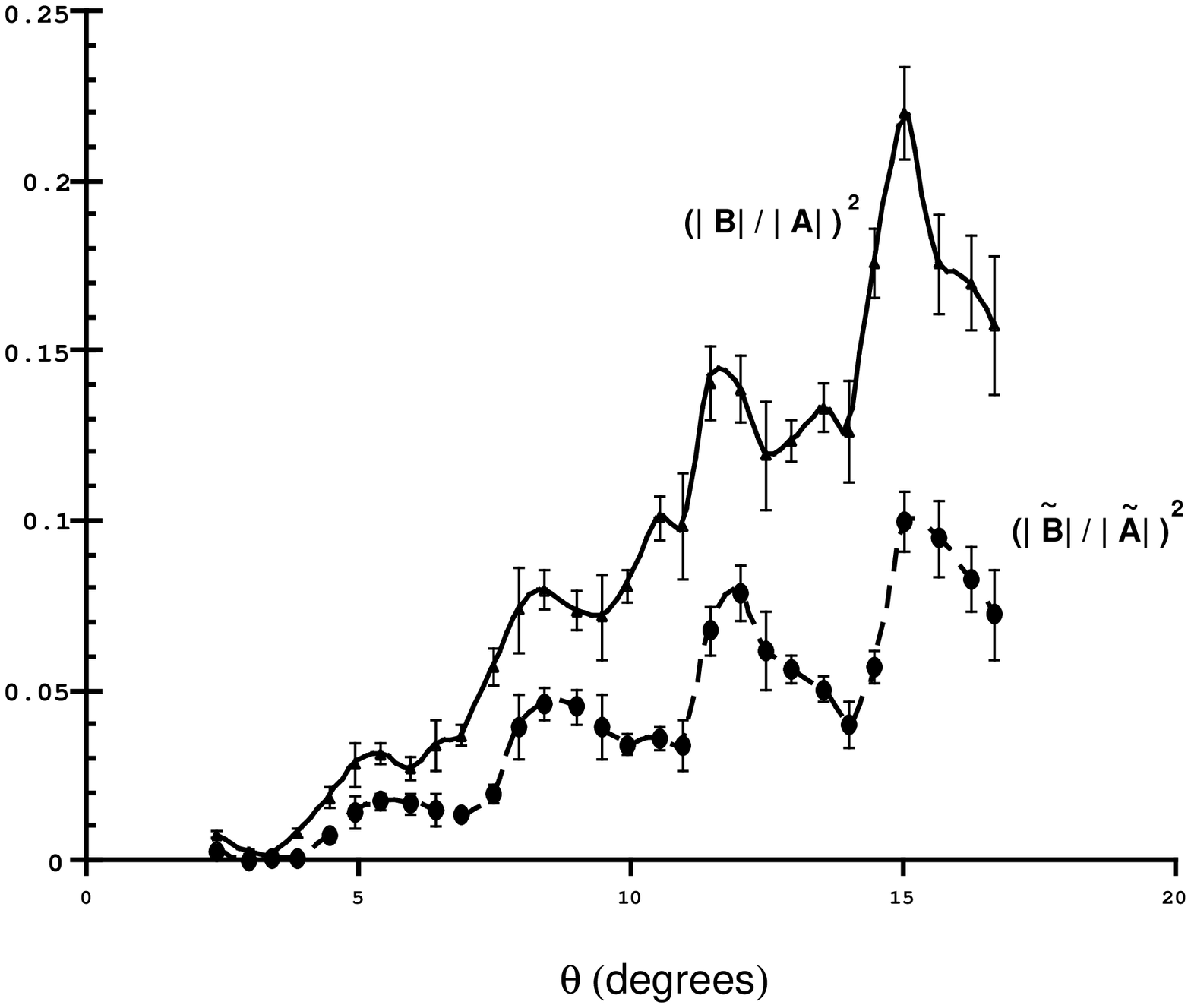 scaled 800}


\begin{thebibliography}{99}

\bibitem {bell}
J. S. Bell and H. Ruegg, {\it Nucl. Phys.} {\bf B98}, 151 (1975). 

\bibitem{gino}

J. N. Ginocchio, {\it Phys. Rev. Lett.} {\bf 78}, 436 (1997). 

\bibitem{gino2}
J. N. Ginocchio and  D. G. Madland, {\it Phys. Rev. C} {\bf 57}, 1167 (1998).

\bibitem{ami} 
J. N. Ginocchio and  A. Leviatan {\it Phys. Lett. B} {\bf 425}, 1 (1998).

\bibitem{gino3} 
J. N. Ginocchio,  {\it Phys. Reports} {\bf 315} (1999); LANL Archives: {\bf nucl-th/9812035}.

\bibitem{wal}
 B. D. Serot and J. D. Walecka, {\it The Relativistic Nuclear Many - Body
Problem} in {\it Advances in Nuclear Physics}, edited by J. W. Negele and
E. Vogt, Vol.\ {\bf 16} (New York, Plenum, 1986).

\bibitem{mad}

B. A. Nikolaus , T. Hoch,  and  D. G. Madland, {\it Phys. Rev.} {\bf C46}, 
1757 (1992).

\bibitem {arima}
J. Meng, K. Sugawara-Tanabe, S. Yamaji, P. Ring,  and A. Arima, {\it Phys. Rev.} {\bf 58}, R628 (1998).

\bibitem {ring}
G. A. Lalazissis, Y. K. Gambhir, J. P. Maharana, C. S. Warke, and P. Ring, {\it Phys. Rev.} {\bf C58}, 
R45 (1998); LANL archives: {\bf nucl-th/9806009}. 

\bibitem{cohen}

T. D. Cohen, R. J. Furnstahl , K . Griegel, and  X. Jin, {\it Prog. in Part. and Nucl. Phys.} {\bf 35}, 221 (1995). 

\bibitem{kth}

K. T. Hecht and A. Adler, {\it Nucl. Phys.} {\bf A137}, 129 (1969).

\bibitem{aa}

A. Arima, M. Harvey, and  K . Shimizu, {\it Phys. Lett.} {\bf 30B}, 517 (1969).

\bibitem{bohr}

A. Bohr, I. Hamamoto, and B. R. Mottelson, {\it Phys. Scr.} {\bf 26}, 267 (1982).

\bibitem {draayer3}
T. Beuschel, A. L. Blokhin, and J. P. Draayer, {\it Nucl. Phys.} {\bf A619}, (1997).
119
\bibitem{dudek}

J. Dudek, W. Nazarewicz, Z. Szymanski, and G. A. Leander, {\it Phys. Rev.
Lett.} {\bf 59}, 1405 (1987).

\bibitem{twin}
W. Nazarewicz, P. J. Twin, P. Fallon, and J. D. Garrett, {\it Phys. Rev.
Lett.} {\bf64}, 1654 (1990).

\bibitem{stephens}
 F. S. Stephens {\it et al}, {\it Phys. Rev.} {\bf C57}, R1565 (1998). 
\bibitem{ben}

B. Mottelson, {\it Nucl. Phys.} {\bf A522}, 1 (1991). 

\bibitem{gino4} 
J. N. Ginocchio, {\it Phys. Rev C} (1999); LANL Archives: {\bf nucl-th/9812025}.

\bibitem{dave} 
 R. Kozack and D. G. Madland, {\it Phys. Rev.} {\bf C39}, 1461 (1989). 

\bibitem{bunny} 
 E. D. Cooper, S. Hama, B. C. Clark, and R. L. Mercer, {\it Phys. Rev.} {\bf C47}, 297 (1993). 

\bibitem{fred} 
 J. B. Bowlin, A. S. Goldhaber, and C. Wilkin, {\it Zeitschrift f{\"u}r Physik A} {\bf 331},83 (1988). 
\bibitem{herman}
H. Feshbach, {\it Theoretical Nuclear Physics - Nuclear Reactions}  (New York, John Wiley and Sons, Inc., 1992).

\bibitem{john} 
 R. W. Fergerson {\it et al}, {\it Phys. Rev.} {\bf C33}, 239 (1986). 

\end{thebibliography}
\end{document}